\documentclass{sf2a-conf2011}
\usepackage{graphicx}
\usepackage{hyperref}
\usepackage[]{natbib}  
\usepackage[]{aeguill}
\usepackage{epstopdf}

\def\BibTeX{{\rm B\kern-.05em{\sc i\kern-.025em b}\kern-.08em
    T\kern-.1667em\lower.7ex\hbox{E}\kern-.125emX}}
\bibpunct{(}{)}{;}{a}{}{,}  

\begin{document}

\TitreGlobal{SF2A 2011}

\title{Modeling the polarization of radio-quiet AGN\\from the optical to the X-ray band}

\runningtitle{SF2A 2011}

\author{F. Marin}\address{Observatoire astronomique de Strasbourg, Section Hautes Energies,
			  11 Rue de l'Universit\'e, 67000 Strasbourg, France\\}

\author{R. W. Goosmann$^1$}

\setcounter{page}{237}

\index{Marin, F.}
\index{Goosmann, R. W.}


\maketitle

\begin{abstract}
A thermal active galactic nucleus (AGN) consist of a powerful, broad-band continuum source that is surrounded by several 
reprocessing media with different geometries and compositions. Here we investigate the expected spectropolarimetric signatures 
in the optical/UV and X-ray wavebands as they arise from the complex radiative coupling between different, axis-symmetric 
AGN media. Using the latest version of the Monte-Carlo radiative transfer code {\sc stokes}, we obtain spectral fluxes, 
polarization percentages, and polarization position angles. In the optical/UV, we assume unpolarized photons coming from 
a compact source that are reprocessed by an optically-thick, dusty torus and by equatorial and polar electron-scattering 
regions. In the X-ray band, we additionally assume a lamp-post geometry with an X-ray source irradiating the accretion 
disk from above. We compare our results for the two wavebands and thereby provide predictions for future X-ray polarimetric 
missions. These predictions can be based on present-day optical/UV spectropolarimetric observations. In particular, we conclude 
that the observed polarization dichotomy in the optical/UV band should extend into the X-ray range.
\end{abstract}



\section{Introduction}

Since the early observations by \citet{Fath1909}, active galactic nuclei (AGN) have been intensively observed at all 
possible wavelengths using ground-based and space telescopes. 
The core of an AGN cannot be resolved by current optical instruments. In addition to that we find that in type-2 objects 
(those with narrow optical emission lines) the central engine is hidden by optically thick dust blocking most of the light. 
According to the unified scheme of AGN \citep{Antonucci1993}, the obscuring dust is distributed anisotropically and in type-1 
AGN, which show broad optical emission lines, the central region is directly visible. This anisotropic distribution of 
absorbing and scattering media in AGN must induce a net polarization that we can exploit 
in order to investigate the complex radiative coupling between the innermost components of AGN.
In fact, spectropolarimetry is a unique tool to probe the unresolvable parts of AGN thanks to two more independent observables 
it adds: the percentage and the position angle of polarization. So far, spectropolarimetry observations could be performed 
from the radio to the optical/UV band, but with the launch of the GEMS satellite \citep{Swank2011} the first X-ray polarization 
data of bright AGN is soon going to be in reach.

To interpret the data, polarization modeling of the radiative interplay between different AGN components is necessary. 
Such modeling has been conducted previously by a number of authors \citep[see e.g.][]{Kartje1995,Smith2004,Wolf1999,Goosmann2011}. 
For computational reasons, some models are restricted to a single-scattering approach following the suggestion 
by \citet{Henney1995} about the predominance of first-order scattering in optically thick media. But one should bear 
in mind that this argument does not necessarily hold for the multiple scattering between several non-absorbing, 
electron-scattering components. Also, previous modeling is most often limited to a given waveband. 

In this research note, we present a composite, multiple-scattering and reprocessing model of AGN from which spectropolarimetric 
fluxes are computed simultaneously in the optical/UV and in the X-ray band. Our model setup is based on the classical, 
axis-symmetric unified scheme of AGN \citep{Urry1995} and we are particularly interested in the polarization properties 
as a function of wavelength and viewing direction. 
When observing the optical polarization of AGN, a dichotomy is found for the polarization angle \citep{Antonucci1983}: 
at type-2 viewing angles, the position angle of the polarization is most often directed perpendicularly to the central 
radio structure; at type-1 viewing angles, the polarization vector favors a direction that is aligned with the (projected) 
radio axis. Assuming that the radio structure is stretched along the symmetry axis of the torus, our modeling allows us to 
test if we can reproduce this observed dichotomy.
In the following, we define {\it parallel} (or {\it perpendicular}) polarization according to the preferentially observed 
polarization angle of {\it type-1} (or {\it type-2}) AGN. When plotting our results, we distinguish parallel polarization
 by adding a negative sign to the polarization percentage.

\section{Modeling the unified scheme of AGN}
\subsection{Model setup for the AGN structure}

We investigate the radiative coupling between different axis-symmetric emission and reprocessing regions: 
the inner and outer parts of the accretion disk, the obscuring equatorial dust region, and double-conical 
outflows along the polar direction. 

We consider a compact continuum source of unpolarized photons being emitted isotropically according to a power-law 
$F_{\rm \nu} \propto \nu^{-\alpha}$ with index $\alpha = 1$. For the optical/UV part (1600~\AA~-- 8000~\AA) 
we assume the continuum source to be very compact and quasi point-like. For the X-ray range (1~keV -- 100~keV), 
we adopt a lamp-post geometry and include X-ray reprocessing of the primary radiation by the underlying disk. 
The primary source is located at low height on the disk axis subtending a large solid angle with the disk.

The source region is surrounded by a geometrically and optically thin scattering annulus with a flared shape. 
This radiation-supported wedge plays a major role as it produces a parallel polarization signature in type-1 view 
by electron scattering (see e.g. \citealt{Chand1960}, \citealt{Angel1969}, \citealt{Anto1984}, \citealt{Sunyaev1985}). 
At larger radius an optically thick, elliptical dusty torus surrounds the system. It shares the same symmetry plane 
as the flared disk and is responsible for the optical obscuration at type-2 views. The torus funnel supposedly collimates 
a mildly-ionized, optically thin outflow stretched along the symmetry axis of the system. The polar wind has an hourglass 
shape and is centered on the photon source. Parameters defining the shape and the composition of the three/four reprocessing 
regions are summarized in Table 1. The dust model used for the torus at optical/UV wavelengths is based on a prescription 
for Galactic dust as described in \citet{Goosmann2007}. 
In the X-ray band, we assume neutral reprocessing for the torus and for the accretion disk. Details of this reprocessing 
model can be found in \citet{Goosmann2011}.

\begin{table}[]
  {\footnotesize
   \centering
   \begin{tabular}{|c|c|c|c|}
   \hline
      {\bf irradiated accretion disk}     & {\bf flared disk}            & {\bf dusty torus}                      & {\bf polar outflows}\\
   \hline
      (only present for X-rays)           & $R_{\rm min} = 0.02$ pc        & $R_{\rm min} = 0.1$ pc                  & $R_{\rm min} = 0.3$ pc\\
      $R_{\rm disk} =  0.0004$ pc           & $R_{\rm max} = 0.04$ pc        & $R_{\rm max} = 0.5$ pc                   & $R_{\rm max} = 1.8$ pc\\
      $h_{\rm disk} =  3.25 \times 10^{-7}$ pc & half-opening angle = 20$^\circ$  & half-opening angle = 60$^\circ$            & half-opening angle = 40$^\circ$ \\
      vertical optical depth $>$600          & equat. optical depth = 1        & equat. optical depth = 750                & vertical optical depth = 0.03\\
      neutral reprocessing                 & electron scattering          & Mie scattering/neutral reprocessing & electron scattering\\
   \hline
   \end{tabular}
  }
  \caption{Parameters of the different model components. The accretion disk is only present when modeling the X-ray range. The elevated primary X-ray source is located on the disk axis and subtends a half-angle of 76$^\circ$ with the disk. Note that for the polar outflow, the half-opening angle is measured with respect to the vertical, symmetry axis of the torus, while for the flared-disk the half-opening angle is taken with respect to the equatorial plane.}
  \label{Table1}
\end{table}

\subsection{The radiative transfer code {\sc stokes}}
We apply the latest version of the Monte-Carlo code {\sc stokes} (\url{www.stokes-program.info}). 
For details on the code, consult \citet{Goosmann2007} and \citet{Goosmann2011}. It conducts radiative transfer in 
complex emission and reprocessing environments and includes the treatment of polarization. The calculations include multiple 
scattering, an angle-dependent analysis in 3D, and different dust models. The {\sc stokes} code computes the total flux spectrum, 
the polarization angle, the percentage of polarization, and the polarized flux. The reprocessing physics depends on the 
energy band considered. Electron and Mie scattering are assumed in the optical/UV waveband; Compton scattering and neutral 
reprocessing predominate in the X-ray range. 

We compute the spectropolarimetric flux as a function of wavelength or photon energy at a given polar viewing angle, $i$, 
that is measured with respect to the symmetry axis of the system.

\section{Results}
\begin{figure}[t!]
 \centering
   \begin{tabular}{ll}
      \includegraphics[width=0.48\textwidth,clip]{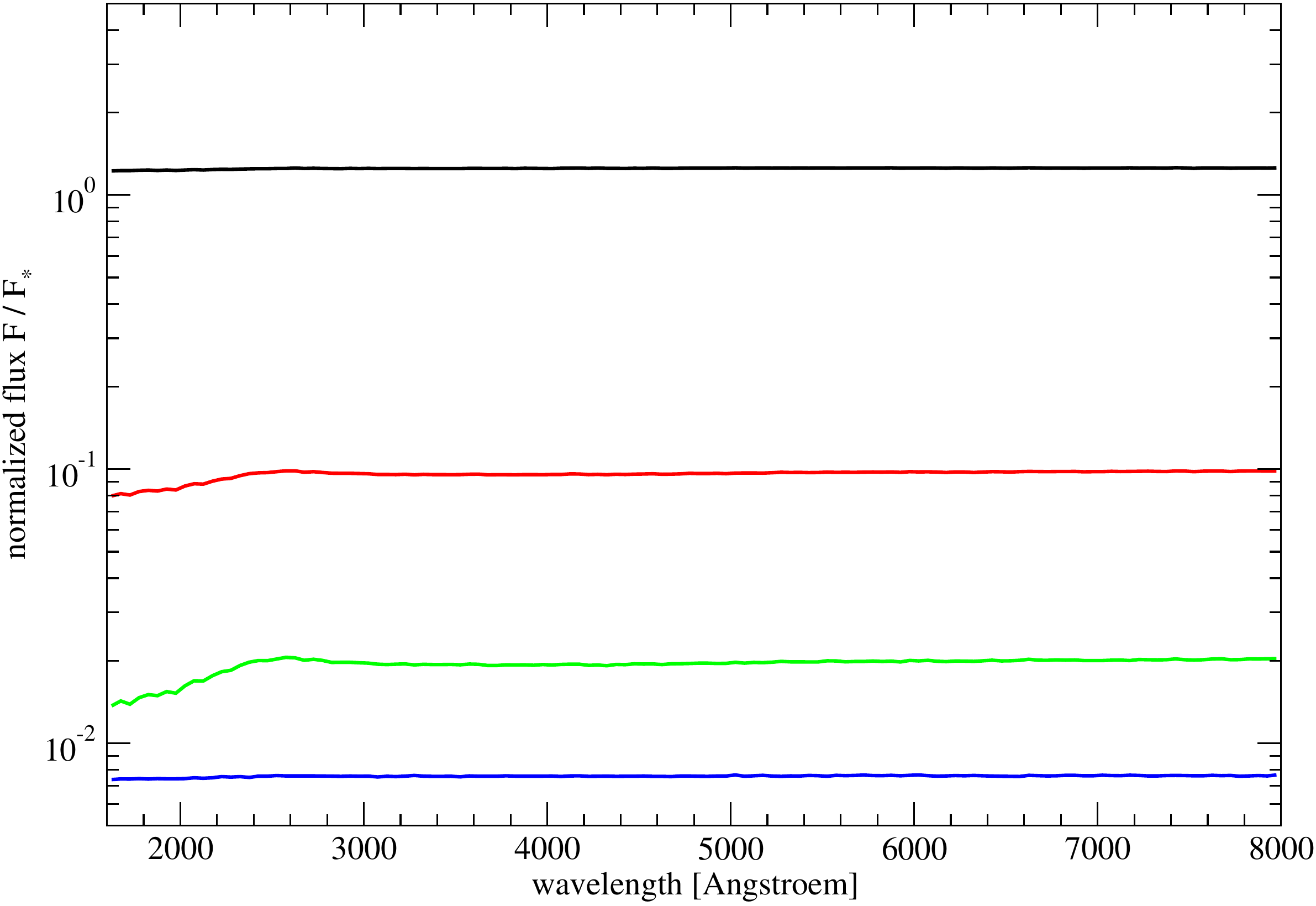} 
      \hspace{0.02\textwidth}
      \includegraphics[width=0.47\textwidth,clip]{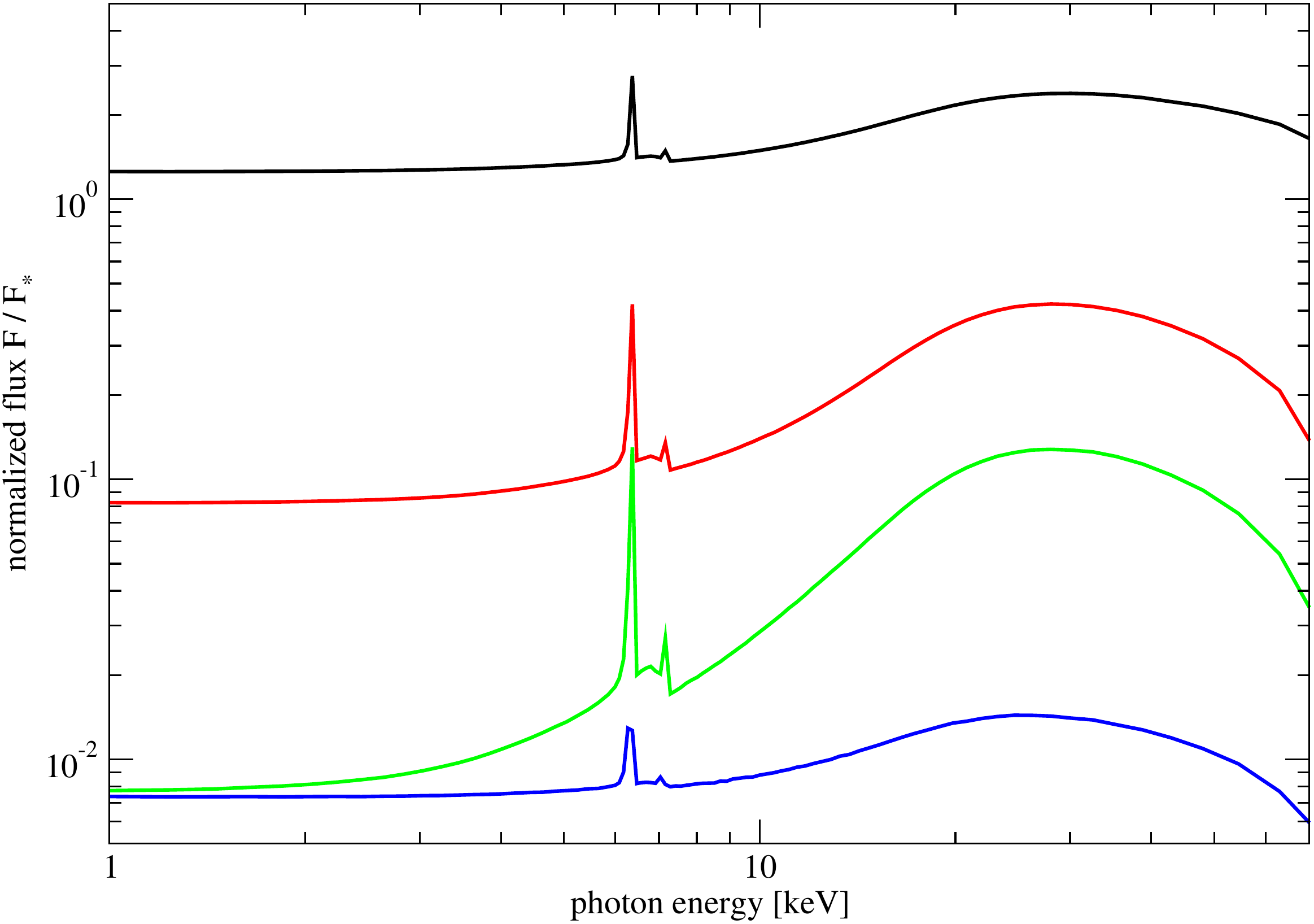} \\
   \end{tabular}
 \centering
   \begin{tabular}{ll}
      \includegraphics[width=0.48\textwidth,clip]{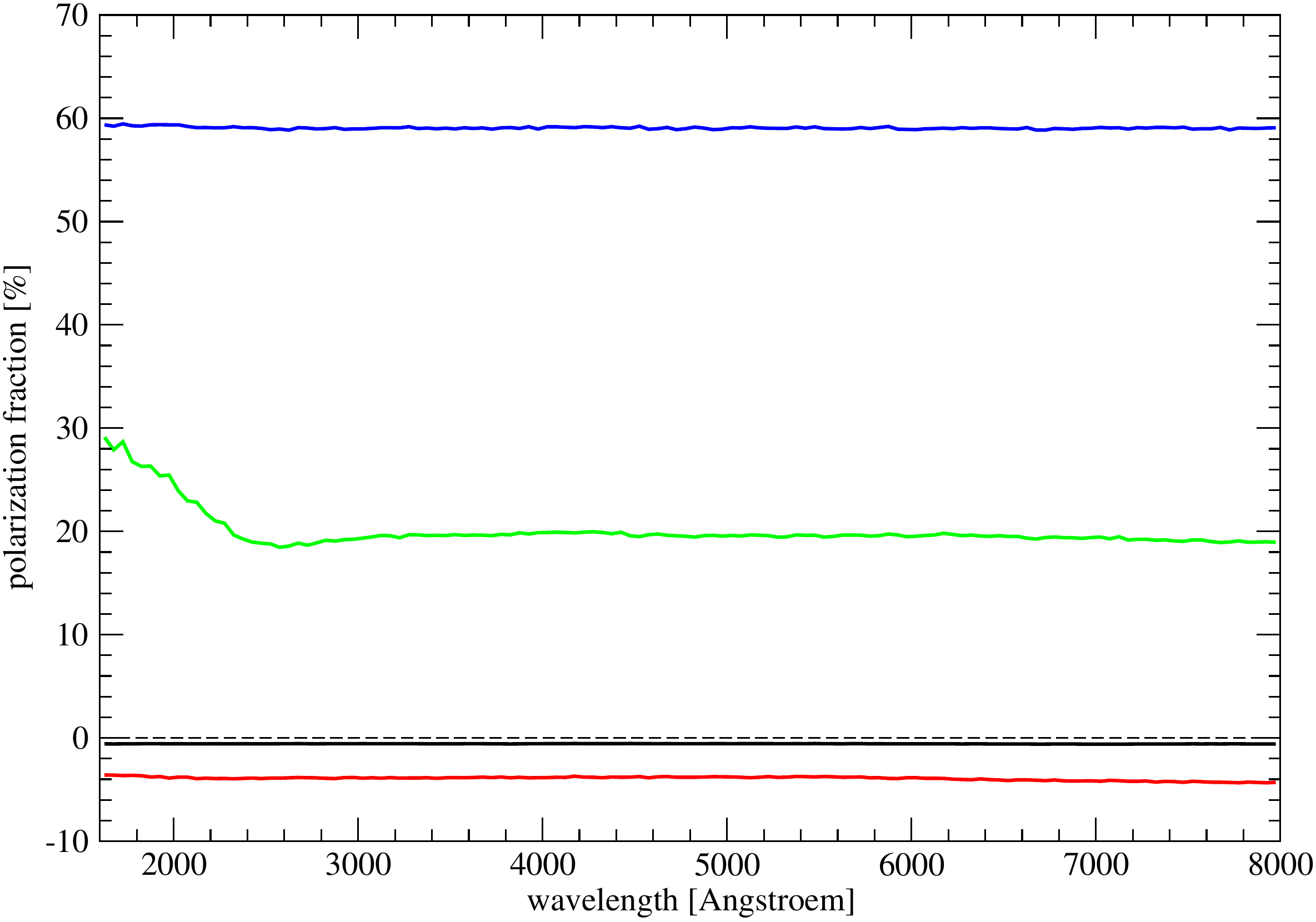} 
      \hspace{0.02\textwidth}
      \includegraphics[width=0.47\textwidth,clip]{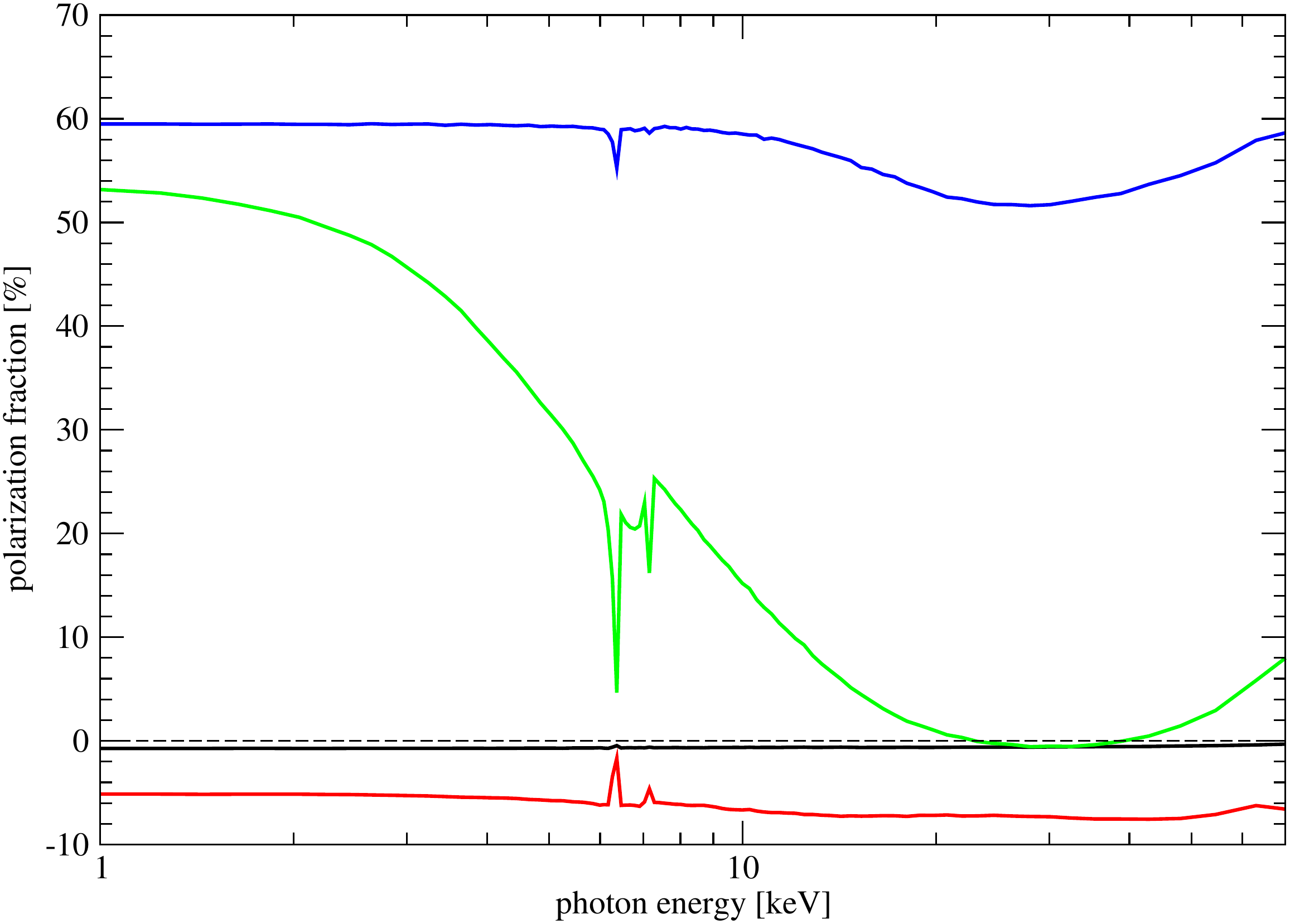} \\
   \end{tabular}
  \caption{Modeling the radiative coupling between the different axis-symmetric reprocessing regions. 
	   The normalized spectral flux is shown in the top panels, the polarization properties are shown below. 
	   A negative $P$ denotes a ``type-1'' polarization (parallel to the projected symmetry axis) and a 
	   positive $P$ stands for a ``type-2'' polarization (perpendicular to the axis). The transition at $P = 0$ is 
	   indicated by dashed lines. Four different viewing angles, $i$, are considered: a face-on view at 
	   $i \sim 18^\circ$ (black), a line-of-sight just below the torus horizon at $\cos i = 0.63^\circ$ (red), 
	   an intermediate type-2 view at $i \sim 76^\circ$ (green), and an edge-on view at $i \sim 87^\circ$ (blue). 
	   Left: optical/UV energy band. Right: X-ray band.}
  \label{author1:fig1}
\end{figure}

In the top panels of Fig.~\ref{author1:fig1}, we present the total flux, $F$, for the two wavebands considered. 
The fluxes are normalized to the pure source flux, $F_*$, that would emerge along the same line-of-sight if there were 
no scattering media. The results for the polarization percentage, $P$, and for the polarization angle are combined in the 
bottom panels of Fig.~\ref{author1:fig1}. We adopt a sign convention for the polarization percentage that is recalled in the figure caption.

\subsection{Results for the optical/UV band}

At face-on and edge-on view, the optical flux is wavelength-independent indicating that electron-scattering in the equatorial flared-disk 
and in the polar outflows dominates. At intermediate viewing-angles, the impact of the wavelength-dependent dust scattering emerges, mostly 
around the $\lambda_{2175}$ feature in the UV. This little bump at 2175~\AA~in the flux spectrum is due to scattering by carbonaceous dust in the torus.

The model reproduces the observed polarization dichotomy. The signature of the flared-disk is visible exceeding the effects of the polar outflow and 
producing low degrees of parallel polarization towards face-on viewing angles. At higher inclination, the equatorial scattering is hidden by the dusty 
torus and polar scattering dominates causing perpendicular polarization. The variations of $P$ with wavelength at intermediate viewing angles indicate 
that the dusty torus also has a significant impact on the polarization. The rise in $P$ towards the UV at $i \sim 73^\circ$ is a combined effect of 
multiple scattering inside the torus funnel and of the wavelength-dependent polarization phase function that is associated with Mie scattering by Galactic dust.

\subsection{Results for the X-ray band}

The X-ray spectrum shows typical features of neutral reprocessing -- the iron K$\alpha$ and K$\beta$ fluorescence lines at 6.4~keV and 7.1~keV 
and their absorption edges, the Compton hump, and strong soft X-ray absorption at intermediate viewing angles are prominent spectral features. 
The fluorescent line emission and the Compton reflection hump around 30 keV are present at every line of sight.

An important result of our modeling is that we predict a polarization dichotomy also for the X-ray band. At all viewing directions below the 
torus horizon, $P$ is positive implying perpendicular polarization. But towards a face-on view, the electron scattering in the equatorial flared 
disk predominates and produces a net parallel polarization. A peculiar feature appears at $i \sim 73^\circ$, where $P$ changes from positive to 
mildly negative values around the Compton hump. This behavior is due to the competition between parallel and perpendicular polarization emerging 
from different reprocessing regions of the model. Around 30~keV, the effect of the flared disk becomes less important than the Compton scattering 
in the other regions. Explaining this behavior in detail is not trivial as several factors have to be taken into account, one of them being the 
angle-dependent scattering phase function. But also,the energy-dependence of the electron scattering cross-section has an effect as it favors 
soft X-ray photons to scatter more than hard X-ray photons. Higher energy photons therefore pass more easily through the optically thin, 
equatorial scattering region without interacting. This partly explains the disappearance of the parallel polarization at higher photon energy. 
A more detailed discussion about the X-ray polarization signature of isolated and coupled reprocessing regions is going to be provided elsewhere.

\section{Summary and conclusions}

We have applied the latest version of the {\sc stokes} radiative transfer code to examine the complex reprocessing between different, 
axis-symmetric media of an active nucleus. We provide simultaneous results for the optical/UV and for the X-ray wavebands and we trace the spectral 
flux and the polarization as a function of photon wavelength. The observed optical/UV polarization dichotomy is successfully reproduced and an 
analogous dichotomy is predicted for the X-ray range.

This work is carried out in anticipation of the forthcoming age of X-ray polarimetry. The NASA space telescope GEMS \citep{Swank2011} is 
planned to be launched in 2014 and will be entirely dedicated to X-ray polarimetry. The satellite will observe X-ray sources in the 2--10~keV 
band allowing us to test the soft X-ray part of our modeling results for the brightest AGN. Note that a next generation, broad-band X-ray 
polarimeter is technically feasible \citep{Tagliaferri2011} and could even observe polarization up to 35~keV. Such observations include the X-ray 
polarization of the Compton hump and thus put even stronger constraints on the validity of our modeling results.

\begin{acknowledgements}
The authors are grateful to Martin Gaskell at the University of Valpara\'iso in Chile for his great help.
\end{acknowledgements}

\bibliographystyle{aa} 
\bibliography{marin} 

\end{document}